\documentclass[copyright]{eptcs}
\providecommand{\event}{FMAS 2023}

\usepackage[T1]{fontenc}
\usepackage[utf8]{inputenc}

\usepackage{amsmath}
\usepackage{amssymb}
\usepackage{xfrac} 
\usepackage[cal=euler]{mathalfa} 
\usepackage[hang]{footmisc}

\usepackage{microtype}

\usepackage[inline]{enumitem}

\usepackage{booktabs}
\usepackage{multirow}
\usepackage{graphicx}
\usepackage{subcaption}

\usepackage{float}

\usepackage{tikz}
\usetikzlibrary{external, positioning}
\usepackage{pgfplots}
\usepackage{pgfplotstable}
\usepgfplotslibrary{colormaps, colorbrewer, groupplots}

\newcommand{\drawDiagonal}{
    \draw[dashed] 
        (axis cs:\pgfkeysvalueof{/pgfplots/xmin},\pgfkeysvalueof{/pgfplots/xmin}) 
        -- 
        (axis cs:\pgfkeysvalueof{/pgfplots/xmax},\pgfkeysvalueof{/pgfplots/xmax});
}

\newcommand{\qty}[2]{#1\,#2}
\newcommand{\num}[1]{#1}
\newcommand{\percent}{\%}
\newcommand{\second}{\text{s}}

\usepackage{bm}
\newcommand{\vect}[1]{\bm{#1}}

\usepackage[capitalise]{cleveref}

\usepackage{oplotsymbl}
\newcommand{\pythsum}[1][0]{\mathbin{\trianglepalinevh_{#1}}}

\usepackage{colonequals}
\newcommand{\eqdef}{\colonequals}

\pgfplotsset{cycle list/Set1}

\pgfplotsset{
  compat=1.17,
  legend style={font=\tiny},
  every axis label/.append style={font=\footnotesize},
  title style={font=\footnotesize},
  tick label style={font=\scriptsize},
  results/.style={
    width=6.4cm,
    ytick={0,0.2,...,1.0},
    xtick={0,50,...,200},
    ymin=0,
    ymax=1,
    legend pos=outer north east,
    legend cell align={left},
    enlargelimits=true,
    xlabel=Epoch,
    ylabel=Accuracy,
    every axis plot post/.append style={
      thick,
      table/col sep=comma,
      table/x=Epoch,
    }
  },
  group/results/.style={
    group style={
      group size=2 by 1, 
      ylabels at=edge left,
      x descriptions at=edge bottom,
      horizontal sep=0.25cm,
    }, 
    results,
  },
  implications/.style={
    xlabel={$x$},
    ylabel={$y$},
    width=3.4cm,
    height=3.6cm,
    title style={xshift=4ex, yshift=-1.75ex},
    xlabel style={xshift=.5ex, yshift=1.75ex},
    ylabel style={xshift=-.5ex, yshift=1.75ex},
    unbounded coords=jump,
    xtick={0,1},
    ytick={0,1},
    ztick={0,1},
    view={35}{35},
    colormap/Spectral,
    samples=25,
    domain=0:1
  },
  beta-search/.style={
        xlabel={Prediction Accuracy},
        ylabel={Constraint Accuracy},
        scatter,
        every axis label/.append style={font=\footnotesize},
        title style={font=\footnotesize},
        tick label style={font=\scriptsize},
        point meta=explicit symbolic,
        nodes near coords*={%
            \pgfmathparse{\pgfplotspointmeta==0 ? "Baseline" : \pgfplotspointmeta}%
            \pgfmathresult%
        },
        nodes near coords style={font=\tiny, black, anchor=north},
        /pgfplots/.cd,
        every axis plot post/.append style={
            visualization depends on={\thisrow{Beta} \as \pgfplotspointmeta},
            table/col sep=comma, 
            table/x=P-Acc, 
            table/y=C-Acc,
            table/meta=Beta
        }
    }
}

\title{Comparing Differentiable Logics for Learning Systems:\\A Research Preview\thanks{This publication has emanated from research conducted with the financial support of Science Foundation Ireland under Grant number 20/FFP-P/8853.}}
\author{Thomas Flinkow
  \institute{Department of Computer Science\\
    Maynooth University\\
    Maynooth, Ireland}
  \email{thomas.flinkow@mu.ie}
  \and
  Barak A. Pearlmutter\qquad\qquad Rosemary Monahan
  \institute{Department of Computer Science and Hamilton Institute\\
    Maynooth University\\
    Maynooth, Ireland}
  \email{\quad barak@pearlmutter.net \quad\qquad rosemary.monahan@mu.ie}
}
\def\titlerunning{Comparing Differentiable Logics for Learning Systems: A Research Preview}
\def\authorrunning{T. Flinkow, R. Monahan \& B. A. Pearlmutter}

\hypersetup{
  bookmarksnumbered,
  pdftitle    = {\titlerunning},
  pdfauthor   = {\authorrunning},
}

\begin{document}
\maketitle

\begin{abstract}
Extensive research on formal verification of machine learning (ML) systems indicates that learning from data alone often fails to capture underlying background knowledge.
A variety of verifiers have been developed to ensure that a machine-learnt model satisfies correctness and safety properties, however, these verifiers typically assume a trained network with fixed weights.
ML-enabled autonomous systems are required to not only detect incorrect predictions, but should also possess the ability to self-correct, continuously improving and adapting.
A promising approach for creating ML models that inherently satisfy constraints is to encode background knowledge as logical constraints that guide the learning process via so-called differentiable logics.
In this research preview, we compare and evaluate various logics from the literature in weakly-supervised contexts, presenting our findings and highlighting open problems for future work.
Our experimental results are broadly consistent with results reported previously in literature; however, learning with differentiable logics introduces a new hyperparameter that is difficult to tune and has significant influence on the effectiveness of the logics.
\end{abstract}

\section{Introduction}
Advancements in machine learning (ML) in the past few years indicate great potential for applying ML to various domains.
Autonomous systems are one such application domain, but using ML components in such a safety-critical domain presents unique new challenges for formal verification. These include
\begin{enumerate*}[label=(\arabic*)]
  \item ML failing to learn background knowledge from data alone \cite{wangEfficientFormalSafety2018a},
  \item neural networks being susceptible to adversarial inputs, and
  \item a lack of specifications, generally and especially when continuous learning is permitted \cite{farrellExploringRequirementsSoftware}.
\end{enumerate*}
Addressing these challenges is even more important and more difficult when the ML-enabled autonomous system is permitted to continue to learn after deployment, either to adapt to changing environments or to correct and improve itself when errors are detected \cite{chengContinuousSafetyVerification2021a}.

A multitude of neural network verifiers (c.f. \cite{mullerThirdInternationalVerification2022}) have been presented in the past few years; prominent examples include Reluplex~\cite{katzReluplexEfficientSMT2017a}, NNV~\cite{tranNNVNeuralNetwork2020c}, and MN-BaB~\cite{ferrariCompleteVerificationMultiNeuron2022a}.
These solvers use techniques such as satisfiability and reachability analysis, and can verify a variety of properties.
However, these verifiers typically assume trained networks with fixed weights and do not target the learning process itself \cite{kwiatkowskaSafetyVerificationDeep2019}.
One step in the direction of correct-by-construction neural networks are so-called \emph{differentiable logics}, which transform a logical constraint $\phi$ into an additional logical loss term $\mathcal{L}_\text{L}(\phi)$ to minimise when learning, where the logical loss of a constraint $\phi$ is combined with standard cross-entropy loss as $\mathcal{L}=\mathcal{L}_\text{CE}+\lambda\mathcal{L}_\text{L}(\phi)$.
In order to translate logical constraints into loss terms, a mapping must be defined that allows for real-valued truth values, and that is differentiable almost everywhere for use with standard gradient-based methods.
In the following, we give a brief overview of two of these mappings (so-called \emph{differentiable logics}) from popular literature, namely DL2 and fuzzy logics.

\subsection{DL2}
DL2 (``Deep Learning with Differentiable Logics'') \cite{fischerDL2TrainingQuerying2019} is a system for querying and training neural networks with logic. It maps absolute truth to $0$ and other degrees of truth to positive values up to $\infty$ based on the following elementary translation rules:
\begin{equation}\label{eq:dl2}
  \mathcal{L}_\text{DL2}(x\le y)\eqdef\max(x-y, 0),\quad\mathcal{L}_\text{DL2}(x\wedge y)\eqdef\mathcal{L}_\text{DL2}(x)+\mathcal{L}_\text{DL2}(y),\quad\mathcal{L}_\text{DL2}(x\vee y)\eqdef\mathcal{L}_\text{DL2}(x)\cdot\mathcal{L}_\text{DL2}(y).
\end{equation}
Additionally, there is $\mathcal{L}_\text{DL2}(x\neq y)\eqdef\xi[x=y]$, where $\xi>0$ denotes a constant that was found to not have significant influence~\cite{fischerDL2TrainingQuerying2019}, and $[\cdot]$ being the indicator (in Knuth's notation).
From these, other rules such as $\mathcal{L}_\text{DL2}(x<y)\eqdef\mathcal{L}_\text{DL2}(x\le y \wedge x\neq y)$ can be derived.
Negation is handled by pushing the negation inwards to the level of comparison, e.g. $\mathcal{L}_\text{DL2}(\lnot(x\le y))\eqdef\mathcal{L}_\text{DL2}(y<x)$.
DL2 does not have a separate translation for implication, instead translating implication as $\mathcal{L}_\text{DL2}(x\rightarrow y)\eqdef\mathcal{L}_\text{DL2}(\lnot x\vee y)$.
\vspace{-3pt} 
\subsection{Fuzzy Logics}
\begin{table}
  \caption{The t-norms, t-conorms and implications used in our experiments.}
  \label{tab:differentiable_logics_comparison}
  \footnotesize
  \centerline{
  \begin{tabular}{llll}
    \toprule
    \textbf{Name}                     & \textbf{T-norm (Conjunction)}                          & \textbf{T-conorm (Disjunction)}           & \textbf{Implication}                                                               \\ \midrule
    \multirow{2}{*}{Gödel}   & \multirow{2}{*}{$T_\text{G}(x,y)=\min(x,y)$}                  & \multirow{2}{*}{$S_\text{G}(x,y)=\max(x,y)$}     & $I_\text{G}(x,y)=\begin{cases}1,&\text{if $x<y$}\\y\end{cases}$             \\ \addlinespace
                             &                                               &                                  & $I_\text{KD}(x,y)=\max(\overline{x},y)$                                            \\ \midrule
    \L ukasiewicz            & $T_\text{\L K}(x,y)=\max(0, x+y-1)$                              & $S_\text{\L K}(x,y)=\min(1, x+y)$                   & $I_\text{\L K}(x,y)=\min(\overline{x}+y,1)$                                        \\ \midrule
    Yager                    & $T_\text{YG}(x,y)=\max(0, \overline{\overline{x}\pythsum[p]\overline{y}})$ & $S_\text{YG}(x,y)=\min(1, x \pythsum[p] y)$ & $I_\text{YG}(x,y)=\begin{cases}1,&\text{if $x=y=0$}\\y^{\,x}\end{cases}$        \\ \midrule
    \multirow{2}{*}{Product} & \multirow{2}{*}{$T_\text{P}(x,y)=xy$}                         & \multirow{2}{*}{$S_\text{PS}(x,y)=x+y-xy$}        & $I_\text{GG}(x,y)=\begin{cases}1,&\text{if $x<y$}\\\sfrac{y}{x}\end{cases}$ \\ \addlinespace
                             &                                               &                                  & $I_\text{RC}(x,y)=\overline{x}+xy $                                                \\ \bottomrule
    \multicolumn{4}{@{}l@{}}{where $u \pythsum[p] v \eqdef \sqrt[p\!]{\lvert u \rvert^p + \lvert v \rvert^p}$ is the $p$-norm Pythagorean sum, $p \ge 1$, and $\overline{u} \eqdef 1-u$}
  \end{tabular}
  }
\end{table}
Whereas DL2 was designed specifically for deep learning contexts, fuzzy logics are logical systems that have been studied extensively and happen to be suitable for use as differentiable logics due to their many-valued nature, with operators that are often differentiable almost everywhere.
Fuzzy logics express degrees of truth in the unit interval $[0,1]$, with absolute truth mapped to $1$. We use $\mathcal{L}_\text{L}(\phi)\eqdef 1.0-\mathcal{L}_\text{FL}(\phi)$ for the fuzzy logic loss in our implementation to address the inverse notion of truth. 

Fuzzy logics are based on functions $T:[0,1]^2\to[0,1]$ that are commutative, associative, monotonic, and satisfy $T(1,y)=y$.
These are called triangular norms (abbreviated as \emph{t-norms}) and generalise conjunction.
A t-conorm (also called \emph{s-norm}) generalises disjunction and can be obtained from a t-norm using $S(x,y)=1-T(1-x, 1-y)$.
From a t-conorm $S$ and fuzzy negation $N$, one obtains a so-called $(S,N)$-implication (which generalises material implication) as $I(x,y)\eqdef S(N(x), y)$).
Examples of $(S,N)$-implications are the Kleene-Dienes implication $I_\text{KD}$ and Reichenbach implication $I_\text{RC}$, both with the standard negation $N(x)=1-x$.
Other implications generalise the intuitionistic implication and are called $R$-implications, because they use the t-norm residuum $R(x,y)=\sup\{t\in[0, 1] \mid T(x, t) \le y\}$.
Example $R$-implications are the Gödel implication $I_\text{G}$ and Goguen implication $I_\text{GG}$.
The \L ukasiewicz implication $I_\text{\L K}$ is both an $(S,N)$-implication and an $R$-implication.
Other implications are neither---the Yager implication $I_\text{YG}$, for example, is an $f$-generated implication that is obtained using $f(x)=-\ln x$ in $I(x,y)\eqdef f^{-1}(xf(y))$ (with the understanding that $0\cdot\infty=0$).

Additionally, \cite{vankriekenAnalyzingDifferentiableFuzzy2022} propose \emph{sigmoidal} implications in order to prove the derivatives of the original implication, while preserving its characteristics.
In \cref{eq:implication_sigmoidal}, $\sigma(x)\eqdef 1/(1+\exp(-x))$ denotes the standard sigmoidal function and $s$ is a parameter controlling the steepness.
We use the sigmoidal implication in our experiments with $I_\text{RC}$ and $s=9$, as suggested by \cite{vankriekenAnalyzingDifferentiableFuzzy2022}.
\begin{equation}\label{eq:implication_sigmoidal}
  (I(x,y))_s\eqdef\dfrac{(1+\exp(\sfrac{s}{2})) \sigma(sI(x,y)-\sfrac{s}{2})-1}{\exp(\sfrac{s}{2})-1}
\end{equation}
Lastly, given a fuzzy implication $I$ and bijection $\phi:[0,1]^2\to[0,1]$, \cite{baczynskiFuzzyImplications2008} show that the function $(I(x,y))_\phi\eqdef\phi^{-1}I(\phi(x),\phi(y))$ is also a fuzzy implication.
We use this in our experiments with the Reichenbach implication $I_\text{RC}$ and $\phi(x)=x^2$.

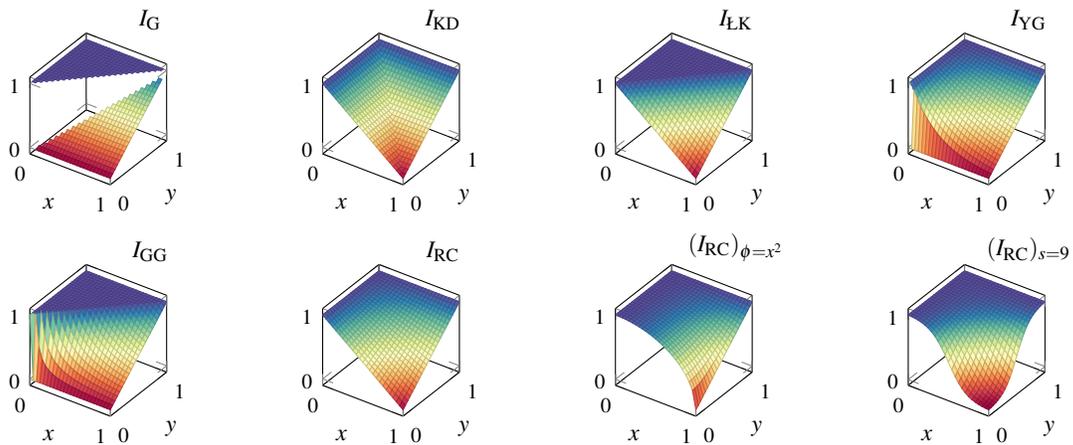
\begin{figure}[H]
  \centering
  \begin{subfigure}[t]{0.2\textwidth}
    \begin{tikzpicture}
      \begin{axis}[implications, title={$I_\text{G}$}]
        \addplot3[surf] {ifthenelse(x<=y, 1, nan)};
        \addplot3[surf] {ifthenelse(x<=y, nan, y)};
      \end{axis}
    \end{tikzpicture}
  \end{subfigure}
  \hspace{0.5cm}
  \begin{subfigure}[t]{0.2\textwidth}
    \begin{tikzpicture}
      \begin{axis}[implications, title={$I_\text{KD}$}]
        \addplot3[surf] {max(1-x, y)};
      \end{axis}
    \end{tikzpicture}
  \end{subfigure}
  \hspace{0.5cm}
  \begin{subfigure}[t]{0.2\textwidth}
    \begin{tikzpicture}
      \begin{axis}[implications, title={$I_\text{\L K}$}]
        \addplot3[surf] {min(1-x+y, 1)};
      \end{axis}
    \end{tikzpicture}
  \end{subfigure}
  \hspace{0.5cm}
  \begin{subfigure}[t]{0.2\textwidth}
    \begin{tikzpicture}
      \begin{axis}[implications, title={$I_\text{YG}$}]
        \addplot3[surf] {ifthenelse(x==0 && y==0, 1, nan)};
        \addplot3[surf] {ifthenelse(x==0 && y==0, nan, y^x)};
      \end{axis}
    \end{tikzpicture}
  \end{subfigure}
  \hspace{0.5cm}
  \begin{subfigure}[t]{0.2\textwidth}
    \begin{tikzpicture}
      \begin{axis}[implications, title={$I_\text{GG}$}]
        \addplot3[surf] {ifthenelse(x<=y, 1, y/x)};
      \end{axis}
    \end{tikzpicture}
  \end{subfigure}
  \hspace{0.5cm}
  \begin{subfigure}[t]{0.2\textwidth}
    \begin{tikzpicture}
      \begin{axis}[implications, title={$I_\text{RC}$}]
        \addplot3[surf] {1-x+x*y};
      \end{axis}
    \end{tikzpicture}
  \end{subfigure}
  \hspace{0.5cm}
  \begin{subfigure}[t]{0.2\textwidth}
    \begin{tikzpicture}
      \begin{axis}[implications, title={$(I_\text{RC})_{\phi=x^2}$}]
        \addplot3[surf] {sqrt(1-x^2+x^2*y^2)};
      \end{axis}
    \end{tikzpicture}
  \end{subfigure}
  \hspace{0.5cm}
  \begin{subfigure}[t]{0.2\textwidth}
    \begin{tikzpicture}[declare function={sigmoid(\x)=1/(1+exp(-\x));IRC(\x,\y)=1-x+x*y;}]
      \begin{axis}[implications, title={$(I_\text{RC})_{s=9}$}]
        \addplot3[surf] {(1+e^(9/2)*sigmoid(9*IRC(x,y)-9/2)-1)/(e^(9/2)-1)};
      \end{axis}
    \end{tikzpicture}
  \end{subfigure}
  \caption{The fuzzy logic implications $I(x,y)$ used in our experiments to map the logical statement $x\rightarrow y$ into real-valued loss. Formal definitions are collected in \cref{tab:differentiable_logics_comparison}.}
  \label{fig:implications}
\end{figure}

\Cref{tab:differentiable_logics_comparison} lists the definitions of the mentioned t-norms, t-conorms and implications, and \cref{fig:implications} (\cref{sec:appendix}) displays plots of the implications.

\paragraph{Mapping atomic terms}
DL2 is designed for atomic terms that are inequalities or comparisons.
As seen in \cref{eq:dl2}, it provides the loss translation $\mathcal{L}_\text{DL2}(x\le y)\eqdef\max(x-y, 0)$ for comparison.

Fuzzy logics typically do not define fuzzy comparison operators. However,
\cite{slusarzLogicDifferentiableLogics2023a} introduce a mapping $\mathcal{L}_\text{FL}(x\le y)\eqdef 1-\max\left(\frac{x-y}{x+y}, 0\right)$ for fuzzy logics.
Fuzzy logics requires the truth values of the atomic terms $x,y$ to be mapped into $[0,1]$ by some oracle.
Because the atomic terms in our constraints are comparisons, we change this mapping from $\mathcal{L}_\text{FL}(x\le y):[0,1]^2\to [0,1]$ to $\mathcal{L}_\text{FL}(x\le y):\mathbb{R}^2\to [0,1]$, allowing us to forgo the need for an external oracle.
The mapping is shown in \cref{eq:oracle_modified} below, where we use $\epsilon=0.05$.
\begin{equation}\label{eq:oracle_modified}
  \mathcal{L}_\text{FL}(x\le y)\eqdef 1-\frac{\max(x-y, 0)}{|x|+|y|+\epsilon}
\end{equation}
Note that the fuzzy logic mapping $\mathcal{L}_\text{FL}(x\le y)$ has a property we intuitively might wish to hold:
for example, we might want $21\leq 20$ to be as much of a violation as $\num{21000}\leq\num{20000}$.
This cannot be achieved in DL2, where the violation depends only on the absolute difference.

\section{Comparing Differentiable Logics: Experimental Setup}
Our comparison experiment\footnote{Available on \url{https://github.com/tflinkow/dl-comparison}.} is implemented in PyTorch and based on the original experiment in \cite{fischerDL2TrainingQuerying2019}.
We train on the Fashion-MNIST, CIFAR-10, and GTSRB data sets with various constraints.
In order to create meaningful scenarios where learning with logical constraints surpasses the baseline (learning from data alone), we train with a fraction of the data sets, namely \qty{10}{\percent} for Fashion-MNIST, \qty{50}{\percent} for CIFAR-10, and \qty{90}{\percent} for GTSRB (as it consists of more classes, with more imperfect data).
Additionally, we introduce \qty{10}{\percent} label noise (training with incorrect labels) for all data sets, and apply various image manipulation techniques, such as random cropping, flipping, and colour changes.
A batch size of 256 was used for all datasets.

The goal of our experiment is to compare various differentiable logics, including DL2 and popular fuzzy logics, and investigate which logic performs most favourable, focusing specifically on implication and conjunction, as these have noticeable consequences for the learning process:
As pointed out by \cite{vankriekenAnalyzingDifferentiableFuzzy2022}, background knowledge and constraints are most often of the form ``if $A$, then $B$''.
Choosing a suitable implication that performs well is thus an important task to guarantee best learning.

\noindent 
In \cite{varnaiRobustnessMetricsLearning2020a}, the authors introduce the \emph{shadow-lifting} property for a conjunction, which requires the truth value of a conjunction to increase when the truth value of a conjunct increases.
This property seems highly desirable for learning, as it allows for gradual improvement.
For example, the formula $0.1\land 1.0$ should be more true than $0.1\land 0.2$, but the Gödel t-norm $T_\text{G}(x,y)=\min(x,y)$ would yield the same truth value in both cases.
DL2 uses addition for conjunction, trivially satisfying shadow-lifting.
The only t-norm to satisfy the shadow-lifting property is the product t-norm $T_\text{P}(x,y)=xy$.
However, as noted by \cite{vankriekenAnalyzingDifferentiableFuzzy2022}, its derivative will be low if $x$ and $y$ are both low, making it hard for the learning process to make progress.

\subsection{Constraints}
\paragraph{Universal quantification}
In \cite{fischerDL2TrainingQuerying2019}, the authors categorise constraints into two distinct schemes: training set constraints, which relate sampled inputs $\vect{x}$ and $\vect{x}'$ from the training set, and global constraints, which concern inputs in the $\epsilon$-ball around a particular input.
They use a PGD-based approach for universally quantified constraints and are thus limited to robustness properties, as noted by \cite{slusarzLogicDifferentiableLogics2023a}.
In \cite{vankriekenAnalyzingDifferentiableFuzzy2022}, the authors relax infinite quantifiers by assuming minibatches to be subsets of an independent distribution and using finite conjunction for universal quantifiers for the minibatch, thus losing soundness.
\cite{slusarzLogicDifferentiableLogics2023a} provide a semantics for quantifiers, independent of the concrete differentiable logic and going beyond robustness via expectation minimisation of a probability distribution.

We do not consider global constraints in our experiment and only utilise finite universal quantification via repeated application of conjunction.
This is permitted due to t-norms being associative and commutative by definition, and DL2's use of addition for conjunction.

\vspace{-3pt} 
\paragraph{Investigating implication}
Limited to only training set constraints, the original DL2 experiment has shown some constraints to already be satisfied in the baseline experiments, where learning with logics would only provide minor improvements.
Their robustness constraint, for example, was already 94.5\% satisfied on Fashion-MNIST for the baseline, compared to 98.36\% with DL2.\footnote{
The only training set constraint where DL2 could significantly improve constraint accuracy in the original experiments (from $5.62\%$ to $99.78\%$ in Fashion-MNIST) is the Lipschitz constraint $\text{Lipschitz}(\mathcal{N}, \vect{x}, \vect{x}', L)\eqdef\lVert\mathcal{N}(\vect{x})-\mathcal{N}(\vect{x}')\rVert_2 \le L\lVert \vect{x}-\vect{x}'\rVert_2$.
We do not include this constraint in our experiments as it does not use conjunction nor implication.
}

They also use a class-similarity constraint for the CIFAR-10 network to encode domain knowledge such as ``a car is more similar to a truck than to a dog''.
In their fully-supervised experiment, even the baseline experiment satisfied this constraint quite well, and DL2 was able to only improve constraint accuracy from $93.67\%$ to $99.68\%$.
As all fuzzy implications $I(x,y)$ by definition behave the same for $x=0$ or $x=1$, the original constraint\footnote{%
\begin{equation*}
  \text{CSim}(\mathcal{N}, \vect{x}, \text{Labels})\eqdef\bigwedge_{(l_1,l_2,l_3)\in \text{Labels}}\; \big((\arg\!\max\mathcal{N}(\vect{x})=l_1)\longrightarrow (\mathcal{N}(\vect{x})_{l_2}\ge \mathcal{N}(\vect{x})_{l_3}) \big).
\end{equation*}
} is not suitable to compare different implications.
Our modified constraint is shown in \cref{eq:csimilarity-t}\footnote{The definition of $\text{Labels}$ is shown in \cref{eq:labels_csimilarity_fmnist} for Fashion-MNIST, and in \cref{eq:labels_csimilarity_cifar10} for CIFAR-10, both in \cref{sec:appendix}.} and replaces the binary decision by a soft one, checking whether the network output for label $l$ is greater than or equal to $\sfrac{1}{\text{\#classes}}$.
\begin{equation}\label{eq:csimilarity-t}
  \text{CSim}(\mathcal{N}, \vect{x}, \text{Labels})\eqdef\bigwedge_{(l_1,l_2,l_3)\in \text{Labels}}\; \big((\mathcal{N}(\vect{x})_{l_1}\ge\sfrac{1}{10})\longrightarrow (\mathcal{N}(\vect{x})_{l_2}\ge \mathcal{N}(\vect{x})_{l_3}) \big).
\end{equation}
As explained above, we translate the conjunction using repeated application of the product t-norm for all fuzzy logics in order to only investigate the different mappings for implication.

\paragraph{Investigating conjunction}
For investigating the shadow-lifting effect, we utilise the German traffic sign recognition benchmark (GTSRB) dataset and use a property that forces the network to make confident, strong decisions by requiring all elements of a group of classes to be either very likely or very unlikely.
Groups consist of classes of a similar type (e.g. speed limit signs)\footnote{The definition of the set of sets $\text{Groups}$ is shown in \cref{eq:groups_gtsrb} (\cref{sec:appendix}).}.
\begin{equation}\label{eq:group-t}
  \text{Group}(\mathcal{N}, \vect{x}, \varepsilon, \text{Groups})\eqdef\bigwedge_{\{g_i\}\in \text{Groups}}\; \big(\sum_{i} \mathcal{N}(\vect{x})_{g_i}\le \varepsilon \;\vee\; \sum_{i} \mathcal{N}(\vect{x})_{g_i} \ge 1 - \varepsilon \big).
\end{equation}
We use the probabilistic sum t-conorm $S_\text{PS}$ for fuzzy disjunctions in order to only focus on the conjunction.

\section{Results}
\begin{figure}[H]
  \centering
  \begin{tikzpicture}
    \begin{groupplot}[group/results, group style={
      group size=2 by 2,
      x descriptions at=edge bottom,
      ylabels at=edge left,
      horizontal sep=0.25cm,
      vertical sep=0.25cm,
    }]
      \nextgroupplot[title=Prediction, ylabel=Accuracy]
        \addplot+[densely dotted] table [y=Test-P-Acc] {reports/report_fmnist_baseline.csv};
        \addplot table [y=Test-P-Acc] {reports/report_fmnist_dl2.csv};
        \addplot table [y=Test-P-Acc] {reports/report_fmnist_g.csv};
        \addplot table [y=Test-P-Acc] {reports/report_fmnist_kd.csv};
        \addplot table [y=Test-P-Acc] {reports/report_fmnist_lk.csv};
        \addplot table [y=Test-P-Acc] {reports/report_fmnist_gg.csv};
        \addplot table [y=Test-P-Acc] {reports/report_fmnist_rc.csv};
        \addplot table [y=Test-P-Acc] {reports/report_fmnist_rc-s.csv};
        \addplot table [y=Test-P-Acc] {reports/report_fmnist_rc-phi.csv};
        \addplot table [y=Test-P-Acc] {reports/report_fmnist_yg.csv};
      \coordinate (c1) at (rel axis cs:0,1);
      \nextgroupplot[title=Constraint, yticklabel pos=right]
        \addplot+[densely dotted] table [y=Test-C-Acc] {reports/report_fmnist_baseline.csv};
        \addplot table [y=Test-C-Acc] {reports/report_fmnist_dl2.csv};
        \addplot table [y=Test-C-Acc] {reports/report_fmnist_g.csv};
        \addplot table [y=Test-C-Acc] {reports/report_fmnist_kd.csv};
        \addplot table [y=Test-C-Acc] {reports/report_fmnist_lk.csv};
        \addplot table [y=Test-C-Acc] {reports/report_fmnist_gg.csv};
        \addplot table [y=Test-C-Acc] {reports/report_fmnist_rc.csv};
        \addplot table [y=Test-C-Acc] {reports/report_fmnist_rc-s.csv};
        \addplot table [y=Test-C-Acc] {reports/report_fmnist_rc-phi.csv};
        \addplot table [y=Test-C-Acc] {reports/report_fmnist_yg.csv};
      \coordinate (c2) at (rel axis cs:1,1);
      
      \nextgroupplot
        \addplot+[densely dotted] table [y=Test-P-Acc] {reports/report_cifar10_baseline.csv};
        \addplot table [y=Test-P-Acc] {reports/report_cifar10_dl2.csv};
        \addplot table [y=Test-P-Acc] {reports/report_cifar10_g.csv};
        \addplot table [y=Test-P-Acc] {reports/report_cifar10_kd.csv};
        \addplot table [y=Test-P-Acc] {reports/report_cifar10_lk.csv};
        \addplot table [y=Test-P-Acc] {reports/report_cifar10_gg.csv};
        \addplot table [y=Test-P-Acc] {reports/report_cifar10_rc.csv};
        \addplot table [y=Test-P-Acc] {reports/report_cifar10_rc-s.csv};
        \addplot table [y=Test-P-Acc] {reports/report_cifar10_rc-phi.csv};
        \addplot table [y=Test-P-Acc] {reports/report_cifar10_yg.csv};
      \coordinate (c1) at (rel axis cs:0,1);
      \nextgroupplot[ 
        yticklabel pos=right, 
        legend style={
          legend columns=5,
          fill=none,
          draw=black,
          anchor=center,
          align=center
        },
        legend to name=full-legend-csim
      ]
        \addplot+[densely dotted] table [y=Test-C-Acc] {reports/report_cifar10_baseline.csv};
        \addplot table [y=Test-C-Acc] {reports/report_cifar10_dl2.csv};
        \addplot table [y=Test-C-Acc] {reports/report_cifar10_g.csv};
        \addplot table [y=Test-C-Acc] {reports/report_cifar10_kd.csv};
        \addplot table [y=Test-C-Acc] {reports/report_cifar10_lk.csv};
        \addplot table [y=Test-C-Acc] {reports/report_cifar10_gg.csv};
        \addplot table [y=Test-C-Acc] {reports/report_cifar10_rc.csv};
        \addplot table [y=Test-C-Acc] {reports/report_cifar10_rc-s.csv};
        \addplot table [y=Test-C-Acc] {reports/report_cifar10_rc-phi.csv};
        \addplot table [y=Test-C-Acc] {reports/report_cifar10_yg.csv};

        \addlegendentry{Baseline}
        \addlegendentry{DL2}
        \addlegendentry{$I_\text{G}$}
        \addlegendentry{$I_\text{KD}$}
        \addlegendentry{$I_\text{\L K}$}
        \addlegendentry{$I_\text{GG}$}
        \addlegendentry{$I_\text{RC}$}
        \addlegendentry{$(I_\text{RC})_{s=9}$}
        \addlegendentry{$(I_\text{RC})_{\phi(x)=x^2}$}
        \addlegendentry{$I_\text{YG}$}
      \coordinate (c2) at (rel axis cs:1,1);
    \end{groupplot}
    \coordinate (c3) at ($(c1)!.5!(c2)$);
    \node[below] at (c3 |- current bounding box.south) {\pgfplotslegendfromname{full-legend-csim}};
  \end{tikzpicture}
  \vspace{-0.25cm}
  \caption{The figure shows how prediction accuracy (left column) and constraint accuracy (right column) change over time when training with the class-similarity constraint for 200 epochs with different logics on Fashion-MNIST (top row) and CIFAR-10 (bottom row).}
  \label{fig:plots_csimilarity}
\end{figure}
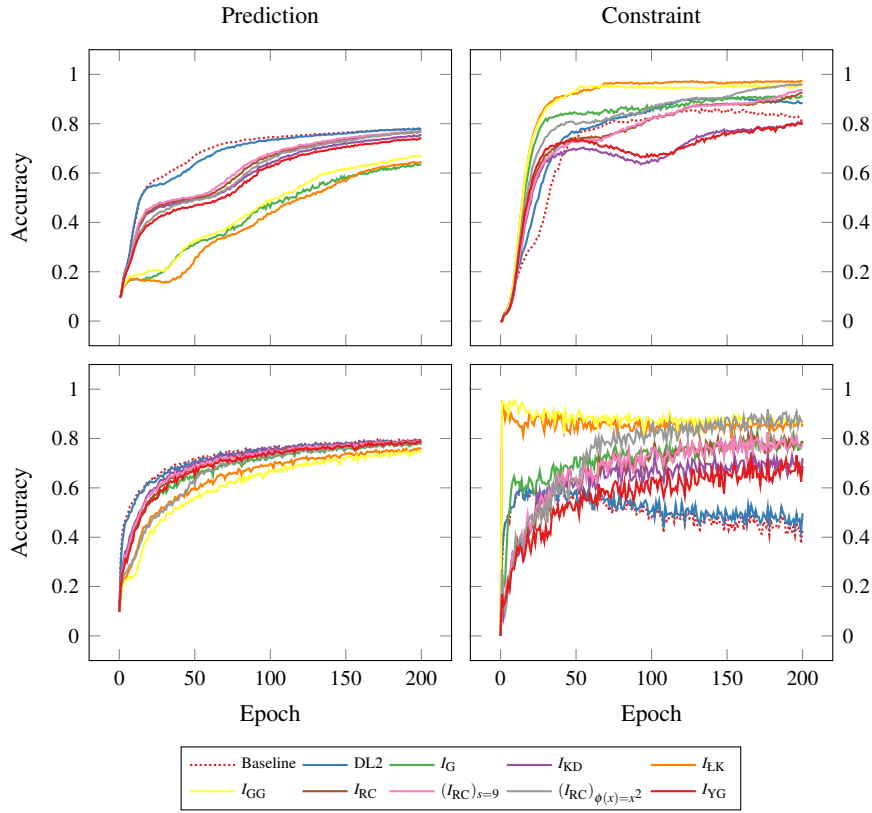

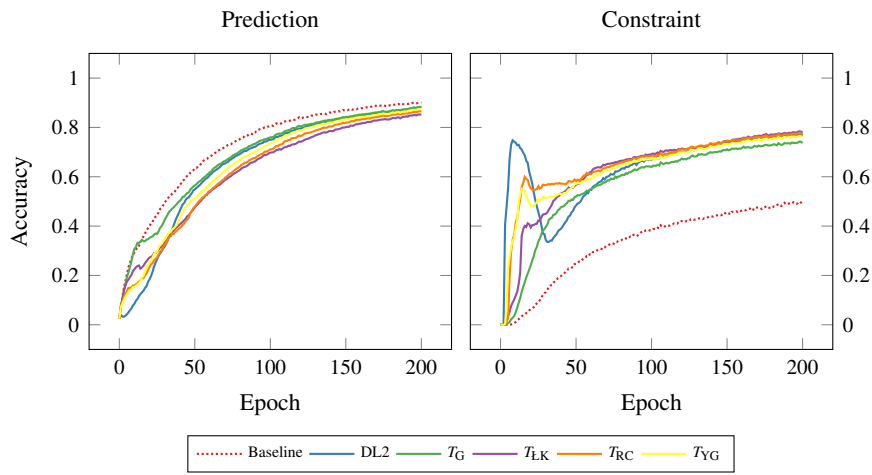
\begin{figure}[H]
  \centering
  \begin{tikzpicture}
    \begin{groupplot}[group/results]
      \nextgroupplot[title=Prediction, ylabel=Accuracy]
        \addplot+[densely dotted] table [y=Test-P-Acc] {reports/report_gtsrb_baseline.csv};
        \addplot table [y=Test-P-Acc] {reports/report_gtsrb_dl2.csv};
        \addplot table [y=Test-P-Acc] {reports/report_gtsrb_g.csv};
        \addplot table [y=Test-P-Acc] {reports/report_gtsrb_lk.csv};
        \addplot table [y=Test-P-Acc] {reports/report_gtsrb_rc.csv};
        \addplot table [y=Test-P-Acc] {reports/report_gtsrb_yg.csv};
      \coordinate (c1) at (rel axis cs:0,1);
      \nextgroupplot[title=Constraint, 
        yticklabel pos=right, 
        legend style={
          at={($(0,0)+(1cm,1cm)$)},
          legend columns=6,
          fill=none,
          draw=black,
          anchor=center,
          align=center
        },
        legend to name=full-legend-gtsrb
      ]
        \addplot+[densely dotted] table [y=Test-C-Acc] {reports/report_gtsrb_baseline.csv};
        \addplot table [y=Test-C-Acc] {reports/report_gtsrb_dl2.csv};
        \addplot table [y=Test-C-Acc] {reports/report_gtsrb_g.csv};
        \addplot table [y=Test-C-Acc] {reports/report_gtsrb_lk.csv};
        \addplot table [y=Test-C-Acc] {reports/report_gtsrb_rc.csv};
        \addplot table [y=Test-C-Acc] {reports/report_gtsrb_yg.csv};

        \addlegendentry{Baseline}
        \addlegendentry{DL2}
        \addlegendentry{$T_\text{G}$}
        \addlegendentry{$T_\text{\L K}$}
        \addlegendentry{$T_\text{RC}$}
        \addlegendentry{$T_\text{YG}$}
      \coordinate (c2) at (rel axis cs:1,1);
    \end{groupplot}
    \coordinate (c3) at ($(c1)!.5!(c2)$);
    \node[below] at (c3 |- current bounding box.south) {\pgfplotslegendfromname{full-legend-gtsrb}};
  \end{tikzpicture}
  \vspace{-0.25cm}
  \caption{The figure shows how prediction accuracy (left column) and constraint accuracy (right column) change over time when training with the group constraint for 200 epochs with different logics on GTSRB.}
  \label{fig:plots_group}
\end{figure}

\Cref{tab:results-csimilarity} shows the results obtained from running the class-similarity constraint experiment on the Fashion-MNIST and CIFAR-10 networks, and \cref{tab:results-group} shows the results obtained from running the group constraint on GTSRB.
For each of these, the displayed prediction and constraint accuracy are obtained by taking the largest of their products from the last 10 epochs.
Additionally, \cref{fig:plots_csimilarity,fig:plots_group} (\cref{sec:appendix}) show how prediction and constraint accuracy change over time.

What immediately stands out is that when training with any logic, constraint accuracy is significantly improved, while prediction accuracy is always slightly reduced, compared to the baseline experiment.
This could be because our constraints fail to capture useful background knowledge that would help with predictions, or, as \cite{fischerDL2TrainingQuerying2019} note, we might observe a similar phenomenon as reported in \cite{tsiprasRobustnessMayBe2018}, where adversarial robustness conflicts with standard generalisation.

Our observed results are broadly consistent with trends previously reported in literature.
The Gödel and Goguen implications perform badly as we expect many wrong inferences (due to the non-existing derivative of $I_\text{G}$ with respect to $x$, and due to the Goguen implication's singularity as $x\to 0$), which manifest in the table as reduced prediction and constraint accuracy.

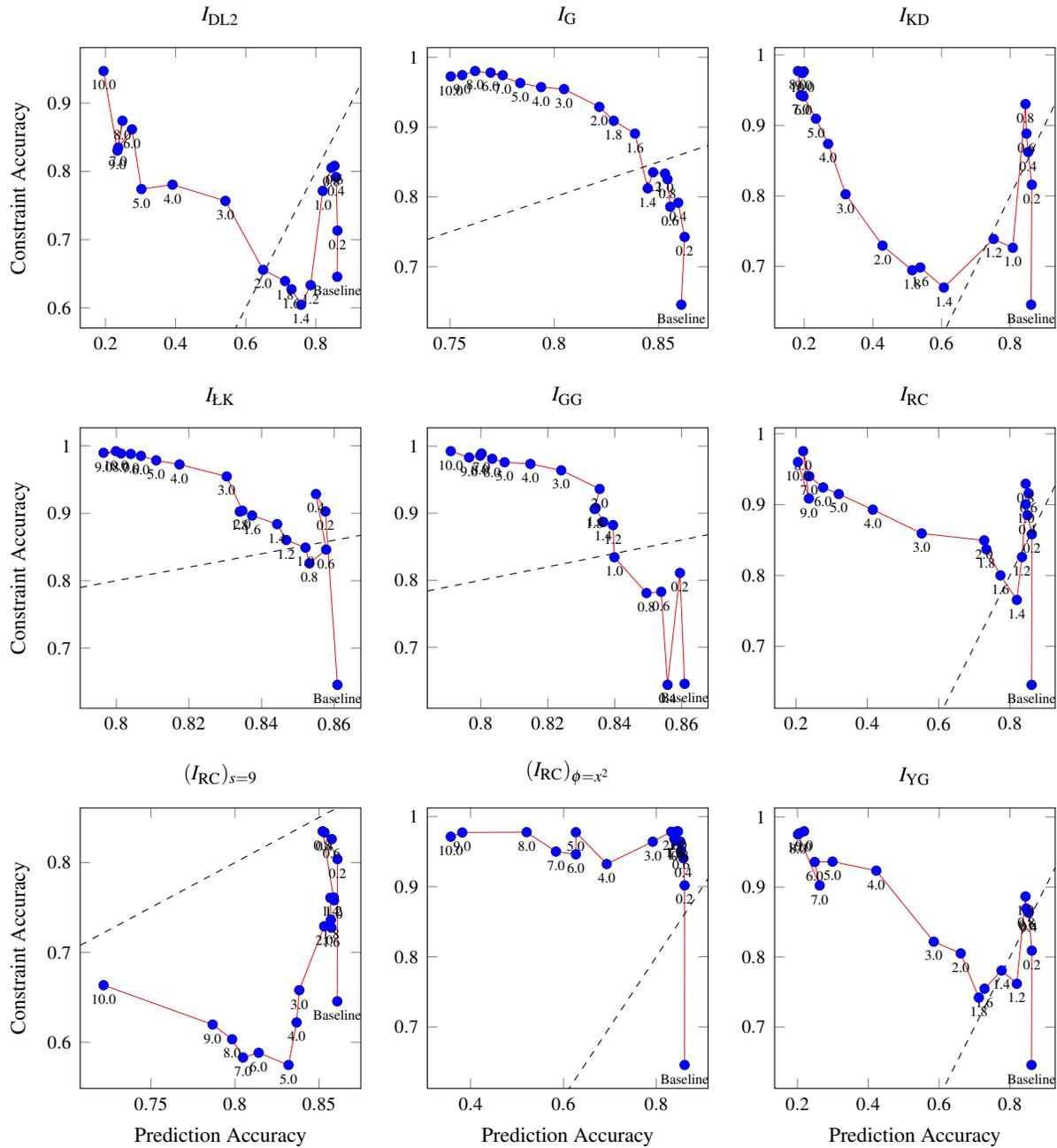
\begin{figure}[H]
  \centering
  \begin{tikzpicture}
    \begin{groupplot}[
      group style={
        group size=3 by 3, 
        ylabels at=edge left,
        xlabels at=edge bottom,
        horizontal sep=1cm,
        vertical sep=1.5cm,
      }, 
      beta-search,
      enlargelimits=true,
      height=5.8cm,
      width=5.8cm]
      \nextgroupplot[title={$I_\text{DL2}$}]
        \addplot table {beta-search/fmnist-dl2.csv};
        \drawDiagonal
      \nextgroupplot[title={$I_\text{G}$}]
        \addplot table {beta-search/fmnist-g.csv};
        \drawDiagonal
      \nextgroupplot[title={$I_\text{KD}$}]
        \addplot table {beta-search/fmnist-kd.csv};
        \drawDiagonal
      \nextgroupplot[title={$I_\text{\L K}$}]
        \addplot table {beta-search/fmnist-lk.csv};
        \drawDiagonal
      \nextgroupplot[title={$I_\text{GG}$}]
        \addplot table {beta-search/fmnist-gg.csv};
        \drawDiagonal
      \nextgroupplot[title={$I_\text{RC}$}]
        \addplot table {beta-search/fmnist-rc.csv};
        \drawDiagonal
      \nextgroupplot[title={$(I_\text{RC})_{s=9}$}]
        \addplot table {beta-search/fmnist-rc-s.csv};
        \drawDiagonal
      \nextgroupplot[title={$(I_\text{RC})_{\phi=x^2}$}]
        \addplot table {beta-search/fmnist-rc-phi.csv};
        \drawDiagonal
      \nextgroupplot[title={$I_\text{YG}$}]
        \addplot table {beta-search/fmnist-yg.csv};
        \drawDiagonal
    \end{groupplot}
  \end{tikzpicture}
  \caption{The figure displays prediction and constraint accuracy obtained when training with varying values of $\lambda$ with the class-similarity constraint on Fashion-MNIST for 200 epochs.}
  \label{fig:plots_beta_search_fmnist}
\end{figure}

\begin{figure}[H]
 \centering
 \begin{tikzpicture}
   \begin{groupplot}[
     group style={
       group size=3 by 3, 
       ylabels at=edge left,
       xlabels at=edge bottom,
       horizontal sep=1cm,
       vertical sep=1.5cm,
     }, 
     beta-search,
     enlargelimits=true,
     height=5.8cm,
     width=5.8cm]
     \nextgroupplot[title={$I_\text{DL2}$}]
       \addplot table {beta-search/cifar10-dl2.csv};
       \drawDiagonal
     \nextgroupplot[title={$I_\text{G}$}]
       \addplot table {beta-search/cifar10-g.csv};
       \drawDiagonal
     \nextgroupplot[title={$I_\text{KD}$}]
       \addplot table {beta-search/cifar10-kd.csv};
       \drawDiagonal
     \nextgroupplot[title={$I_\text{\L K}$}]
       \addplot table {beta-search/cifar10-lk.csv};
       \drawDiagonal
     \nextgroupplot[title={$I_\text{GG}$}]
       \addplot table {beta-search/cifar10-gg.csv};
       \drawDiagonal
     \nextgroupplot[title={$I_\text{RC}$}]
       \addplot table {beta-search/cifar10-rc.csv};
       \drawDiagonal
     \nextgroupplot[title={$(I_\text{RC})_{s=9}$}]
       \addplot table {beta-search/cifar10-rc-s.csv};
       \drawDiagonal
     \nextgroupplot[title={$(I_\text{RC})_{\phi=x^2}$}]
       \addplot table {beta-search/cifar10-rc-phi.csv};
       \drawDiagonal
     \nextgroupplot[title={$I_\text{YG}$}]
       \addplot table {beta-search/cifar10-yg.csv};
       \drawDiagonal
   \end{groupplot}
 \end{tikzpicture}
 \caption{The figure displays prediction and constraint accuracy obtained when training with varying values of $\lambda$ with the class-similarity constraint on CIFAR-10 for 200 epochs.}
 \label{fig:plots_beta_search_cifar10}
\end{figure}
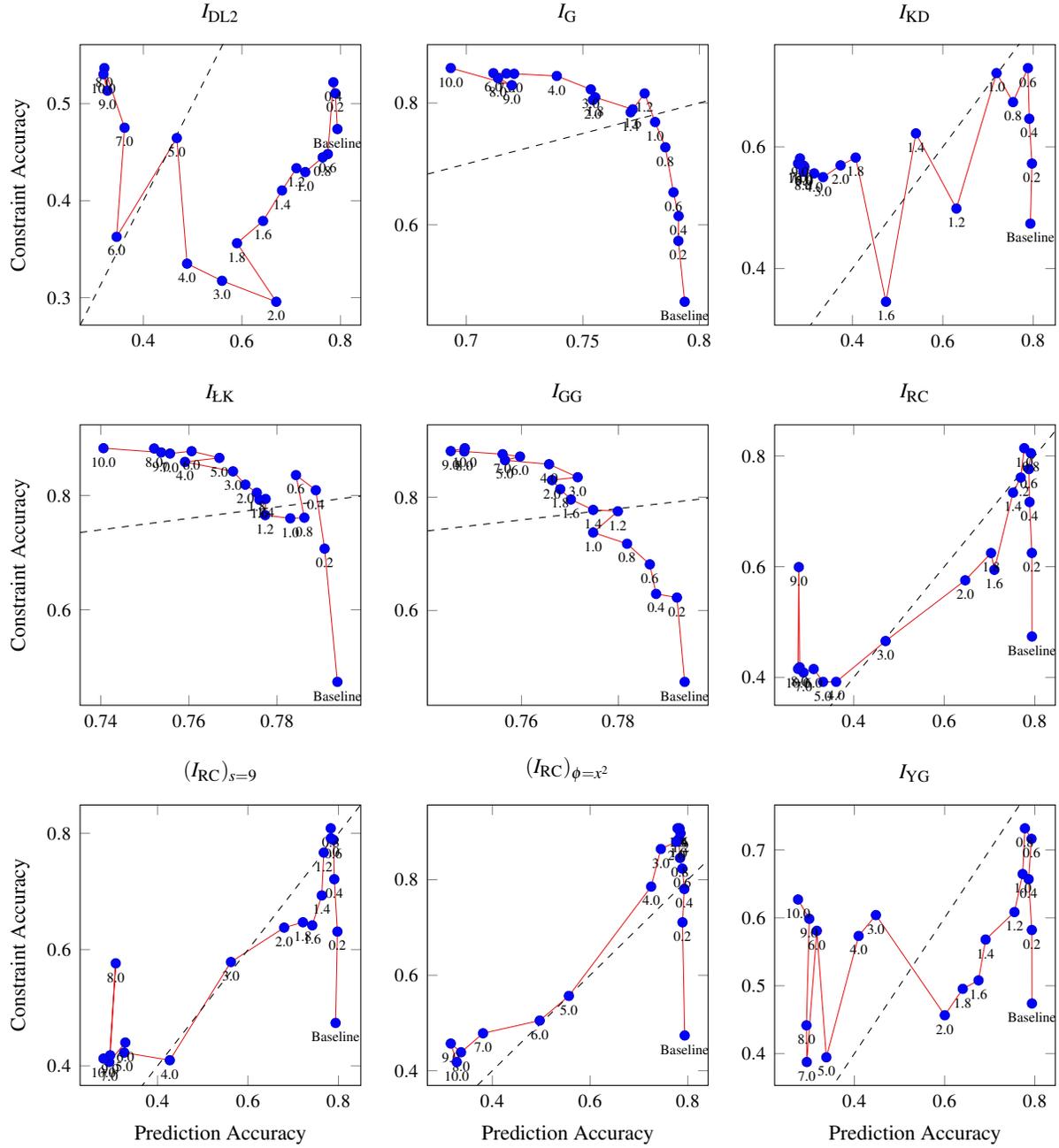

\begin{figure}
 \centering
 \begin{tikzpicture}
   \begin{groupplot}[
     group style={
       group size=3 by 2, 
       ylabels at=edge left,
       xlabels at=edge bottom,
       horizontal sep=1cm,
       vertical sep=1.5cm,
     }, 
     beta-search,
     enlargelimits=true,
     height=5.8cm,
     width=5.8cm]
     \nextgroupplot[title={$I_\text{DL2}$}]
       \addplot table {beta-search/gtsrb-dl2.csv};
       \drawDiagonal
     \nextgroupplot[title={$I_\text{G}$}]
       \addplot table {beta-search/gtsrb-g.csv};
       \drawDiagonal
     \nextgroupplot[title={$I_\text{\L K}$}, ]
       \addplot table {beta-search/gtsrb-lk.csv};
       \drawDiagonal
     \nextgroupplot[title={$I_\text{RC}$}]
       \addplot table {beta-search/gtsrb-rc.csv};
       \drawDiagonal
     \nextgroupplot[title={$I_\text{YG}$}]
       \addplot table {beta-search/gtsrb-yg.csv};
       \drawDiagonal
   \end{groupplot}
 \end{tikzpicture}
 \caption{The figure displays prediction and constraint accuracy obtained when training with varying values of $\lambda$ with the group constraint on GTSRB for 200 epochs.}
 \label{fig:plots_beta_search_gtsrb}
\end{figure}
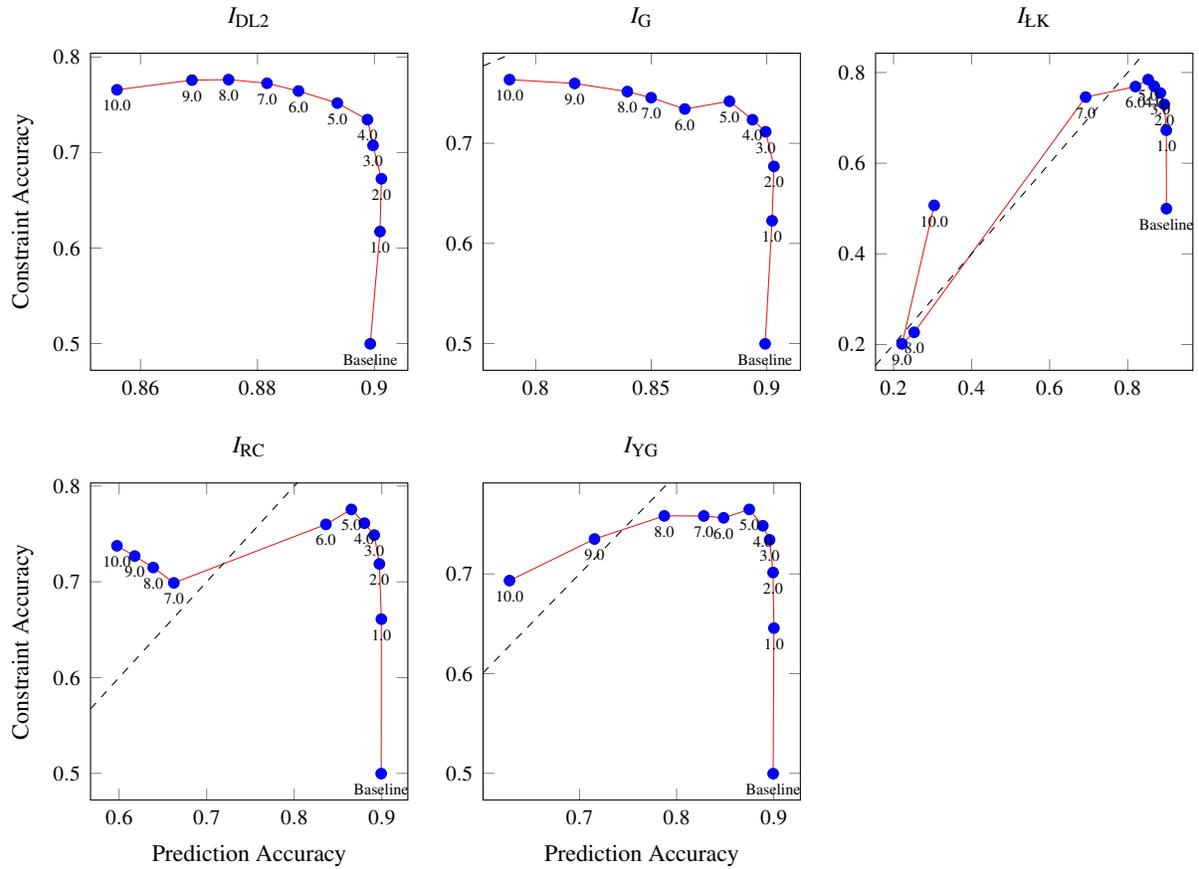

Comparing DL2 and fuzzy logics, our results indicate that for implication, fuzzy logics are the better choice --- for CIFAR-10, even $I_\text{KD}$ performs better than DL2, although both are rewriting $x\rightarrow y$ to $\lnot x\vee y$.
This difference could be due to a multitude of reasons; the mappings $\mathcal{L}(x\le y)$ and $\mathcal{L}(x\wedge y)$ are very different for DL2 and fuzzy logics, as is their range.
Albeit the most likely reason is a sub-optimal choice of hyperparameter $\lambda$, explained in more detail in the next paragraph.
For the group constraint, DL2 performs slightly better than any of the fuzzy logics, although the differences between the logics are overall not as noticeable compared to the class-similarity constraint.
\begin{table}
  \caption{Results. P/C is prediction / constraint accuracy in \%.}
  \begin{subtable}[t]{0.7\textwidth}
    \centering
    \subcaption{Class-Similarity constraint.}
    \label{tab:results-csimilarity}
    \small
    \begin{tabular}{@{}lrrrrrrr@{}}
\toprule
  & \multicolumn{3}{c}{\textbf{Fashion-MNIST}} & \multicolumn{3}{c}{\textbf{CIFAR-10}} \\ \cmidrule(lr){2-4} \cmidrule(lr){5-7}
                                    & \multicolumn{1}{c}{P}               & \multicolumn{1}{c}{C}               & \multicolumn{1}{c}{$\lambda$}          & \multicolumn{1}{c}{P}                & \multicolumn{1}{c}{C}                & \multicolumn{1}{c}{$\lambda$}           \\ \midrule
  Baseline                         & 77.55 & 84.31 & \multicolumn{1}{c}{--}          & 79.06 & 48.65 & \multicolumn{1}{c}{--}           \\
  DL2                              & 77.88 & 89.15      & 0.6    & 78.55 & 52.21      & 0.4    \\
  $I_\text{G}$                  & 63.46 & 91.30        & 3.0      & 77.65 & 81.56        & 1.2      \\
  $I_\text{KD}$                 & 75.39 & 80.59       & 0.8     & 78.82 & 72.94       & 0.6     \\
  $I_\text{\L K}$              & 64.64 & 97.28       & 4.0     & 76.06 & 87.75       & 6.0     \\
  $I_\text{GG}$                 & 67.25 & 95.54       & 3.0     & 74.83 & 88.67       & 10.0     \\
  $I_\text{RC}$                 & 76.79 & 92.56       & 0.8     & 79.14 & 80.51       & 0.8     \\
  $(I_\text{RC})_{s=9}$       & 77.06 & 93.63     & 0.8   & 78.30 & 80.87     & 0.8   \\
  $(I_\text{RC})_{\phi=x^2}$ & \textbf{76.88} & \textbf{95.94}   & 1.0 & \textbf{78.31} & \textbf{90.74}   & 1.6 \\
  $I_\text{YG}$                 & 74.14 & 80.15       & 1.0     & 77.81 & 73.19       & 0.8     \\ \bottomrule
\end{tabular}
  \end{subtable}%
  \begin{subtable}[t]{0.3\textwidth}
    \centering
    \subcaption{Group constraint.}
    \label{tab:results-group}
    \small
    \begin{tabular}{@{}lrrr@{}}
  \toprule
  & \multicolumn{3}{c}{\textbf{GTSRB}} \\ \cmidrule(lr){2-4}
                       & \multicolumn{1}{c}{P}              & \multicolumn{1}{c}{C}              & \multicolumn{1}{c}{$\lambda$}      \\ \midrule
  Baseline            & 89.93 & 49.97 & \multicolumn{1}{c}{--}      \\
  DL2                 & \textbf{88.16} & \textbf{77.25}      & 7.0 \\
  $T_\text{G}$     & 88.38 & 74.20        & 5.0   \\
  $T_\text{\L K}$ & 85.26 & 78.43       & 5.0  \\
  $T_\text{RC}$    & 86.52 & 77.56       & 5.0  \\
  $T_\text{YG}$    & 87.47 & 76.47       & 5.0  \\ \bottomrule
\end{tabular}
  \end{subtable}
\end{table}
\paragraph{Hyperparameter $\lambda$}
Learning with logical loss introduces the hyperparameter $\lambda$, which is the logical weight for the total loss calculation.
Finding a suitable logical weight $\lambda$ is crucial for achieving good results, as choosing a sub-optimal value can potentially result in operators that are supposed to perform badly (such as the Gödel implication) performing even better than logics that are supposed to perform best.
Our strategy to approximate good values of $\lambda$ was to run the same experiment (data set, constraint, logic) for the same number of epochs for each logical weight value $\lambda\in\{0,0.2,0.4,0.6,0.8,1,2,3,4,5,6,7,8,9,10\}$.
We then selected the value that yielded the highest combined prediction and constraint accuracy.
\Cref{tab:results-csimilarity,tab:results-group} show the value of $\lambda$ we chose for each run, and \cref{fig:plots_beta_search_fmnist,fig:plots_beta_search_cifar10,fig:plots_beta_search_gtsrb} (\cref{sec:appendix}) show the prediction and constraint accuracies for each value of $\lambda$. 

This approach is very expensive and not feasible for real-world application.
Unfortunately, extrapolating from running experiments at a smaller number of epochs does not necessarily transfer over to running the experiment at the desired number of epochs.
Further, as can be seen in \cref{fig:plots_beta_search_fmnist,fig:plots_beta_search_cifar10,fig:plots_beta_search_gtsrb} (\cref{sec:appendix}), the resulting graphs are non-monotonic, making it difficult to predict the prediction and constraint accuracy one would obtain with other values of $\lambda$. 

\vspace{-4pt} 
\section{Future Work}
\vspace{-2pt} 
Our experiments have shown that learning with differentiable logics can generally improve how much a ML model satisfies a constraint.
Imposing logical constraints on the training process in this manner could be a step in the direction of verified ML, allowing the use of continuous-learning in self-improving ML-enabled autonomous systems.
It has to be noted that in contrast to formal verifiers, training with logical loss does not formally guarantee properties to hold in all possible cases.

We highlight a few more areas for future work in the following.
\vspace{-4pt} 
\paragraph{Reusing logical constraints during inference}
Because of the differentiable logics acting as a regulariser during training, any logical constraints imposed on the learning process are unavailable during inference.
The trained model can therefore not make use of the logical constraints to check its predictions, for example to attach confidence scores to its predictions.
\vspace{-4pt} 
\paragraph{Probabilistic logics}
Despite expressing satisfaction of formulas on $[0,1]$, fuzzy logics are inherently not probabilistic, having been designed instead for reasoning in the presence of vagueness.
We point to DeepProbLog~\cite{manhaeveDeepProbLogNeuralProbabilistic2018} as one example for a probabilistic logic for use with deep learning.
In the context of neural networks, which often output probabilities, it could be more natural to reason about probabilities instead of vagueness, especially for constraints that include probabilities~\cite{farrellExploringRequirementsSoftware}.
\vspace{-4pt} 
\paragraph{Properties}
A common problem with verifying ML is the lack of specifications, as noted by \cite{leuckerFormalVerificationNeural2020,farrellExploringRequirementsSoftware}.
Most properties in the literature are limited to robustness against slight perturbations, although differentiable logics can not only relate network inputs and outputs, but could also refer to the inner workings of the neural network, such as weights and activation states.
A related area is to investigate whether learning with logical loss can be used to show that desired properties continue to hold when retraining the network.

\bibliographystyle{eptcs}
\bibliography{references}

\begin{thebibliography}{10}
\providecommand{\bibitemdeclare}[2]{}
\providecommand{\surnamestart}{}
\providecommand{\surnameend}{}
\providecommand{\urlprefix}{Available at }
\providecommand{\url}[1]{\texttt{#1}}
\providecommand{\href}[2]{\texttt{#2}}
\providecommand{\urlalt}[2]{\href{#1}{#2}}
\providecommand{\doi}[1]{doi:\urlalt{https://doi.org/#1}{#1}}
\providecommand{\eprint}[1]{arXiv:\urlalt{https://arxiv.org/abs/#1}{#1}}
\providecommand{\bibinfo}[2]{#2}

\bibitemdeclare{book}{baczynskiFuzzyImplications2008}
\bibitem{baczynskiFuzzyImplications2008}
\bibinfo{author}{Micha{\l} \surnamestart Baczy{\'n}ski\surnameend} \&
  \bibinfo{author}{Balasubramaniam \surnamestart Jayaram\surnameend}
  (\bibinfo{year}{2008}): \emph{\bibinfo{title}{Fuzzy Implications}}.
\newblock {\slshape \bibinfo{series}{Studies in Fuzziness and Soft Computing}}
  \bibinfo{volume}{v. 231}, \bibinfo{publisher}{{Springer Verlag}},
  \bibinfo{address}{{Berlin}}, \doi{10.1007/978-3-540-69082-5}.

\bibitemdeclare{inproceedings}{chengContinuousSafetyVerification2021a}
\bibitem{chengContinuousSafetyVerification2021a}
\bibinfo{author}{Chih-Hong \surnamestart Cheng\surnameend} \&
  \bibinfo{author}{Rongjie \surnamestart Yan\surnameend}
  (\bibinfo{year}{2021}): \emph{\bibinfo{title}{Continuous {{Safety
  Verification}} of {{Neural Networks}}}}.
\newblock In: {\slshape \bibinfo{booktitle}{2021 {{Design}}, {{Automation}} \&
  {{Test}} in {{Europe Conference}} \& {{Exhibition}} ({{DATE}})}}, pp.
  \bibinfo{pages}{1478--1483}, \doi{10.23919/DATE51398.2021.9473994}.

\bibitemdeclare{inproceedings}{farrellExploringRequirementsSoftware}
\bibitem{farrellExploringRequirementsSoftware}
\bibinfo{author}{Marie \surnamestart Farrell\surnameend},
  \bibinfo{author}{Anastasia \surnamestart Mavridou\surnameend} \&
  \bibinfo{author}{Johann \surnamestart Schumann\surnameend}
  (\bibinfo{year}{2023}): \emph{\bibinfo{title}{Exploring Requirements for
  Software that Learns: {A} Research Preview}}.
\newblock In \bibinfo{editor}{Alessio \surnamestart Ferrari\surnameend} \&
  \bibinfo{editor}{Birgit \surnamestart Penzenstadler\surnameend}, editors:
  {\slshape \bibinfo{booktitle}{Requirements Engineering: Foundation for
  Software Quality - 29th International Working Conference, {REFSQ} 2023,
  Barcelona, Spain, April 17-20, 2023, Proceedings}}, {\slshape
  \bibinfo{series}{Lecture Notes in Computer Science}} \bibinfo{volume}{13975},
  \bibinfo{publisher}{Springer}, pp. \bibinfo{pages}{179--188},
  \doi{10.1007/978-3-031-29786-1\_12}.

\bibitemdeclare{misc}{ferrariCompleteVerificationMultiNeuron2022a}
\bibitem{ferrariCompleteVerificationMultiNeuron2022a}
\bibinfo{author}{Claudio \surnamestart Ferrari\surnameend},
  \bibinfo{author}{Mark~Niklas \surnamestart Muller\surnameend},
  \bibinfo{author}{Nikola \surnamestart Jovanovic\surnameend} \&
  \bibinfo{author}{Martin \surnamestart Vechev\surnameend}
  (\bibinfo{year}{2022}): \emph{\bibinfo{title}{Complete {{Verification}} via
  {{Multi-Neuron Relaxation Guided Branch-and-Bound}}}},
  \doi{10.48550/arXiv.2205.00263}.

\bibitemdeclare{inproceedings}{fischerDL2TrainingQuerying2019}
\bibitem{fischerDL2TrainingQuerying2019}
\bibinfo{author}{Marc \surnamestart Fischer\surnameend},
  \bibinfo{author}{Mislav \surnamestart Balunovic\surnameend},
  \bibinfo{author}{Dana \surnamestart {Drachsler-Cohen}\surnameend},
  \bibinfo{author}{Timon \surnamestart Gehr\surnameend},
  \bibinfo{author}{Ce~\surnamestart Zhang\surnameend} \&
  \bibinfo{author}{Martin \surnamestart Vechev\surnameend}
  (\bibinfo{year}{2019}): \emph{\bibinfo{title}{{{DL2}}: {{Training}} and
  {{Querying Neural Networks}} with {{Logic}}}}.
\newblock In: {\slshape \bibinfo{booktitle}{Proceedings of the 36th
  {{International Conference}} on {{Machine Learning}}}},
  \bibinfo{publisher}{{PMLR}}, pp. \bibinfo{pages}{1931--1941}.

\bibitemdeclare{inproceedings}{katzReluplexEfficientSMT2017a}
\bibitem{katzReluplexEfficientSMT2017a}
\bibinfo{author}{Guy \surnamestart Katz\surnameend}, \bibinfo{author}{Clark
  \surnamestart Barrett\surnameend}, \bibinfo{author}{David~L. \surnamestart
  Dill\surnameend}, \bibinfo{author}{Kyle \surnamestart Julian\surnameend} \&
  \bibinfo{author}{Mykel~J. \surnamestart Kochenderfer\surnameend}
  (\bibinfo{year}{2017}): \emph{\bibinfo{title}{Reluplex: {{An Efficient SMT
  Solver}} for {{Verifying Deep Neural Networks}}}}.
\newblock In \bibinfo{editor}{Rupak \surnamestart Majumdar\surnameend} \&
  \bibinfo{editor}{Viktor \surnamestart Kun{\v c}ak\surnameend}, editors:
  {\slshape \bibinfo{booktitle}{Computer {{Aided Verification}}}},
  \bibinfo{series}{Lecture {{Notes}} in {{Computer Science}}},
  \bibinfo{publisher}{{Springer International Publishing}},
  \bibinfo{address}{{Cham}}, pp. \bibinfo{pages}{97--117},
  \doi{10.1007/978-3-319-63387-9\_5}.

\bibitemdeclare{inproceedings}{kwiatkowskaSafetyVerificationDeep2019}
\bibitem{kwiatkowskaSafetyVerificationDeep2019}
\bibinfo{author}{M.Z. \surnamestart Kwiatkowska\surnameend}
  (\bibinfo{year}{2019}): \emph{\bibinfo{title}{Safety Verification for Deep
  Neural Networks with Provable Guarantees}}.
\newblock In: {\slshape \bibinfo{booktitle}{Leibniz {{International
  Proceedings}} in {{Informatics}}, {{LIPIcs}}}}, \bibinfo{volume}{140},
  \doi{10.4230/lipics.concur.2019.1}.

\bibitemdeclare{inproceedings}{leuckerFormalVerificationNeural2020}
\bibitem{leuckerFormalVerificationNeural2020}
\bibinfo{author}{Martin \surnamestart Leucker\surnameend}
  (\bibinfo{year}{2020}): \emph{\bibinfo{title}{Formal {{Verification}} of
  {{Neural Networks}}?}}
\newblock In \bibinfo{editor}{Gustavo \surnamestart Carvalho\surnameend} \&
  \bibinfo{editor}{Volker \surnamestart Stolz\surnameend}, editors: {\slshape
  \bibinfo{booktitle}{Formal {{Methods}}: {{Foundations}} and
  {{Applications}}}}, \bibinfo{series}{Lecture {{Notes}} in {{Computer
  Science}}}, \bibinfo{publisher}{{Springer International Publishing}},
  \bibinfo{address}{{Cham}}, pp. \bibinfo{pages}{3--7},
  \doi{10.1007/978-3-030-63882-5\_1}.

\bibitemdeclare{inproceedings}{manhaeveDeepProbLogNeuralProbabilistic2018}
\bibitem{manhaeveDeepProbLogNeuralProbabilistic2018}
\bibinfo{author}{Robin \surnamestart Manhaeve\surnameend},
  \bibinfo{author}{Sebastijan \surnamestart Dumancic\surnameend},
  \bibinfo{author}{Angelika \surnamestart Kimmig\surnameend},
  \bibinfo{author}{Thomas \surnamestart Demeester\surnameend} \&
  \bibinfo{author}{Luc \surnamestart De~Raedt\surnameend}
  (\bibinfo{year}{2018}): \emph{\bibinfo{title}{{{DeepProbLog}}: {{Neural
  Probabilistic Logic Programming}}}}.
\newblock In: {\slshape \bibinfo{booktitle}{Advances in {{Neural Information
  Processing Systems}}}}, \bibinfo{volume}{31}, \bibinfo{publisher}{{Curran
  Associates, Inc.}}

\bibitemdeclare{misc}{mullerThirdInternationalVerification2022}
\bibitem{mullerThirdInternationalVerification2022}
\bibinfo{author}{Mark~Niklas \surnamestart M{\"u}ller\surnameend},
  \bibinfo{author}{Christopher \surnamestart Brix\surnameend},
  \bibinfo{author}{Stanley \surnamestart Bak\surnameend},
  \bibinfo{author}{Changliu \surnamestart Liu\surnameend} \&
  \bibinfo{author}{Taylor~T. \surnamestart Johnson\surnameend}
  (\bibinfo{year}{2022}): \emph{\bibinfo{title}{The {{Third International
  Verification}} of {{Neural Networks Competition}} ({{VNN-COMP}} 2022):
  {{Summary}} and {{Results}}}}, \doi{10.48550/arXiv.2212.10376}.

\bibitemdeclare{inproceedings}{slusarzLogicDifferentiableLogics2023a}
\bibitem{slusarzLogicDifferentiableLogics2023a}
\bibinfo{author}{Natalia \surnamestart {\'S}lusarz\surnameend},
  \bibinfo{author}{Ekaterina \surnamestart Komendantskaya\surnameend},
  \bibinfo{author}{Matthew \surnamestart Daggitt\surnameend},
  \bibinfo{author}{Robert \surnamestart Stewart\surnameend} \&
  \bibinfo{author}{Kathrin \surnamestart Stark\surnameend}
  (\bibinfo{year}{2023}): \emph{\bibinfo{title}{Logic of {{Differentiable
  Logics}}: {{Towards}} a {{Uniform Semantics}} of {{DL}}}}.
\newblock In: {\slshape \bibinfo{booktitle}{{{EPiC Series}} in {{Computing}}}},
  \bibinfo{volume}{94}, \bibinfo{publisher}{{EasyChair}}, pp.
  \bibinfo{pages}{473--493}, \doi{10.29007/c1nt}.

\bibitemdeclare{inproceedings}{tranNNVNeuralNetwork2020c}
\bibitem{tranNNVNeuralNetwork2020c}
\bibinfo{author}{Hoang-Dung \surnamestart Tran\surnameend},
  \bibinfo{author}{Xiaodong \surnamestart Yang\surnameend},
  \bibinfo{author}{Diego \surnamestart Manzanas~Lopez\surnameend},
  \bibinfo{author}{Patrick \surnamestart Musau\surnameend},
  \bibinfo{author}{Luan~Viet \surnamestart Nguyen\surnameend},
  \bibinfo{author}{Weiming \surnamestart Xiang\surnameend},
  \bibinfo{author}{Stanley \surnamestart Bak\surnameend} \&
  \bibinfo{author}{Taylor~T. \surnamestart Johnson\surnameend}
  (\bibinfo{year}{2020}): \emph{\bibinfo{title}{{{NNV}}: {{The Neural Network
  Verification Tool}} for {{Deep Neural Networks}} and {{Learning-Enabled
  Cyber-Physical Systems}}}}.
\newblock In \bibinfo{editor}{Shuvendu~K. \surnamestart Lahiri\surnameend} \&
  \bibinfo{editor}{Chao \surnamestart Wang\surnameend}, editors: {\slshape
  \bibinfo{booktitle}{Computer {{Aided Verification}}}},
  \bibinfo{series}{Lecture {{Notes}} in {{Computer Science}}},
  \bibinfo{publisher}{{Springer International Publishing}},
  \bibinfo{address}{{Cham}}, pp. \bibinfo{pages}{3--17},
  \doi{10.1007/978-3-030-53288-8\_1}.

\bibitemdeclare{inproceedings}{tsiprasRobustnessMayBe2018}
\bibitem{tsiprasRobustnessMayBe2018}
\bibinfo{author}{Dimitris \surnamestart Tsipras\surnameend},
  \bibinfo{author}{Shibani \surnamestart Santurkar\surnameend},
  \bibinfo{author}{Logan \surnamestart Engstrom\surnameend},
  \bibinfo{author}{Alexander \surnamestart Turner\surnameend} \&
  \bibinfo{author}{Aleksander \surnamestart Madry\surnameend}
  (\bibinfo{year}{2018}): \emph{\bibinfo{title}{Robustness {{May Be}} at
  {{Odds}} with {{Accuracy}}}}.
\newblock In: {\slshape \bibinfo{booktitle}{International {{Conference}} on
  {{Learning Representations}}}}.

\bibitemdeclare{article}{vankriekenAnalyzingDifferentiableFuzzy2022}
\bibitem{vankriekenAnalyzingDifferentiableFuzzy2022}
\bibinfo{author}{Emile \surnamestart {van Krieken}\surnameend},
  \bibinfo{author}{Erman \surnamestart Acar\surnameend} \&
  \bibinfo{author}{Frank \surnamestart {van Harmelen}\surnameend}
  (\bibinfo{year}{2022}): \emph{\bibinfo{title}{Analyzing {{Differentiable
  Fuzzy Logic Operators}}}}.
\newblock {\slshape \bibinfo{journal}{Artificial Intelligence}}
  \bibinfo{volume}{302}, p. \bibinfo{pages}{103602},
  \doi{10.1016/j.artint.2021.103602}.
\newblock \eprint{2002.06100}.

\bibitemdeclare{inproceedings}{varnaiRobustnessMetricsLearning2020a}
\bibitem{varnaiRobustnessMetricsLearning2020a}
\bibinfo{author}{Peter \surnamestart Varnai\surnameend} \&
  \bibinfo{author}{Dimos~V. \surnamestart Dimarogonas\surnameend}
  (\bibinfo{year}{2020}): \emph{\bibinfo{title}{On {{Robustness Metrics}} for
  {{Learning STL Tasks}}}}.
\newblock In: {\slshape \bibinfo{booktitle}{2020 {{American Control
  Conference}} ({{ACC}})}}, pp. \bibinfo{pages}{5394--5399},
  \doi{10.23919/ACC45564.2020.9147692}.

\bibitemdeclare{inproceedings}{wangEfficientFormalSafety2018a}
\bibitem{wangEfficientFormalSafety2018a}
\bibinfo{author}{Shiqi \surnamestart Wang\surnameend}, \bibinfo{author}{Kexin
  \surnamestart Pei\surnameend}, \bibinfo{author}{Justin \surnamestart
  Whitehouse\surnameend}, \bibinfo{author}{Junfeng \surnamestart
  Yang\surnameend} \& \bibinfo{author}{Suman \surnamestart Jana\surnameend}
  (\bibinfo{year}{2018}): \emph{\bibinfo{title}{Efficient {{Formal Safety
  Analysis}} of {{Neural Networks}}}}.
\newblock In: {\slshape \bibinfo{booktitle}{Advances in {{Neural Information
  Processing Systems}}}}, \bibinfo{volume}{31}, \bibinfo{publisher}{{Curran
  Associates, Inc.}}

\end{thebibliography}

\appendix
\section{Appendix}\label{sec:appendix}

\subsection{Constraints}
The full set of labels used in the class-similarity constraint for Fashion-MNIST is shown in \cref{eq:labels_csimilarity_fmnist}, the set of labels for CIFAR-10 is shown in \cref{eq:labels_csimilarity_cifar10}.
Note that we have a conjunct for each class so as to rule out cases where the implication would vacuously be true, which does not necessarily reflect real world constraints.

\begin{align}\label{eq:labels_csimilarity_fmnist}
  \text{Labels}_\text{Fashion-MNIST}\eqdef
  \left\{
    \substack{
      (\texttt{T-shirt/top}, \texttt{Shirt}, \texttt{Ankle boot}), \\
      (\texttt{Trouser}, \texttt{Dress}, \texttt{Bag}), \\
      (\texttt{Pullover}, \texttt{Shirt}, \texttt{Sandal}), \\
      (\texttt{Dress}, \texttt{Coat}, \texttt{Bag}), \\
      (\texttt{Coat}, \texttt{Pullover}, \texttt{Shirt}), \\
      (\texttt{Sandal}, \texttt{Sneaker}, \texttt{Dress}), \\
      (\texttt{Shirt}, \texttt{Pullover}, \texttt{Sneaker}), \\
      (\texttt{Sneaker}, \texttt{Sandal}, \texttt{Trouser}), \\
      (\texttt{Bag}, \texttt{Sandal}, \texttt{Dress}), \\
      (\texttt{Ankle boot}, \texttt{Sneaker}, \texttt{T-shirt/top})
    }
  \right\}
\end{align}

\begin{align}\label{eq:labels_csimilarity_cifar10}
  \text{Labels}_\text{CIFAR-10}\eqdef
  \left\{
      \substack{
      (\texttt{airplane}, \texttt{ship}, \texttt{dog}), \\
      (\texttt{automobile}, \texttt{truck}, \texttt{cat}), \\
      (\texttt{bird}, \texttt{airplane}, \texttt{dog}), \\
      (\texttt{cat}, \texttt{dog}, \texttt{frog}), \\
      (\texttt{deer}, \texttt{horse}, \texttt{truck}), \\
      (\texttt{dog}, \texttt{cat}, \texttt{bird}), \\
      (\texttt{frog}, \texttt{ship}, \texttt{truck}), \\
      (\texttt{horse}, \texttt{deer}, \texttt{airplane}), \\
      (\texttt{ship}, \texttt{airplane}, \texttt{deer}), \\
      (\texttt{truck}, \texttt{automobile}, \texttt{airplane})
      }
  \right\}
\end{align}

The definitions of the groups used in the group constraint \cref*{eq:group-t} for GTSRB are given in \cref{eq:groups_gtsrb}.

\begin{gather}
  \text{Group}_\text{Speed Limits}\eqdef
  \left\{
    \substack{
      \texttt{limit 20km/h}, \\
      \texttt{limit 30km/h}, \\
      \texttt{limit 50km/h}, \\
      \texttt{limit 60km/h}, \\
      \texttt{limit 70km/h}, \\
      \texttt{limit 80km/h}, \\
      \texttt{end of limit 80km/h}, \\
      \texttt{limit 100km/h}, \\
      \texttt{limit 120km/h}
    }
  \right\},
  \qquad\text{Group}_\text{Mandatory Actions}\eqdef
  \left\{
    \substack{
      \texttt{turn right ahead}, \\
      \texttt{turn left ahead}, \\
      \texttt{ahead only}, \\
      \texttt{go straight or right}, \\
      \texttt{go straight or left}, \\
      \texttt{keep right}, \\
      \texttt{keep left}, \\
      \texttt{roundabout}
    }
  \right\}
  \nonumber
  \\
  \text{Group}_\text{Prohibitions}\eqdef
  \left\{
    \substack{
      \texttt{no passing}, \\
      \texttt{no passing for trucks}, \\
      \texttt{no way}, \\
      \texttt{no way one-way}, \\
      \texttt{end of no passing}, \\
      \texttt{end of no passing for trucks}
    }
  \right\},
  \qquad\text{Group}_\text{Warnings}\eqdef
  \left\{
    \substack{
      \texttt{caution general}, \\
      \texttt{caution curve left}, \\
      \texttt{caution curve right}, \\
      \texttt{caution curvy}, \\
      \texttt{caution bumps}, \\
      \texttt{caution slippery}, \\
      \texttt{caution narrow road}, \\
      \texttt{road work}, \\
      \texttt{pedestrians}, \\
      \texttt{children crossing}, \\
      \texttt{wild animals crossing}
    }
  \right\}
  \nonumber
  \\
  \text{Groups}_\text{GTSRB}\eqdef\{\text{Group}_\text{Speed Limits}, \text{Group}_\text{Prohibitions}, \text{Group}_\text{Mandatory Actions}, \text{Group}_\text{Warnings} \}
  \label{eq:groups_gtsrb}
\end{gather}

\newpage
\tikzexternaldisable
\subsection{Plots of Prediction and Constraint Accuracy Over Time}

\subsection{Runtime Overhead}
\begin{table}[H]
  \caption{Average epoch train times in seconds.}
  \label{tab:runtimes}
  \begin{subtable}[t]{0.7\textwidth}
    \centering
    \subcaption{Class-Similarity constraint.}
    \label{tab:results-csimilarity-times}
    \small
    \begin{tabular}{@{}lrr@{}}
  \toprule
                                    & \textbf{Fashion-MNIST}                       & \textbf{CIFAR-10}                             \\ \midrule
  Baseline                         & \qty{0.6}{\second} & \qty{3.4}{\second} \\
  DL2                              & \qty{0.6}{\second}      & \qty{3.4}{\second}      \\
  $I_\text{G}$                  & \qty{0.6}{\second}        & \qty{3.8}{\second}        \\
  $I_\text{KD}$                 & \qty{0.7}{\second}       & \qty{4.0}{\second}       \\
  $I_\text{\L K}$              & \qty{0.7}{\second}       & \qty{4.0}{\second}       \\
  $I_\text{GG}$                 & \qty{0.7}{\second}       & \qty{4.0}{\second}       \\
  $I_\text{RC}$                 & \qty{0.7}{\second}       & \qty{4.0}{\second}       \\
  $(I_\text{RC})_{s=9}$       & \qty{0.7}{\second}     & \qty{4.1}{\second}     \\
  $(I_\text{RC})_{\phi=x^2}$ & \qty{0.7}{\second}   & \qty{4.1}{\second}   \\
  $I_\text{YG}$                 & \qty{0.7}{\second}       & \qty{4.1}{\second}       \\ \bottomrule
\end{tabular}
  \end{subtable}%
  \begin{subtable}[t]{0.3\textwidth}
    \centering
    \subcaption{Group constraint.}
    \label{tab:results-group-times}
    \small
    \begin{tabular}{@{}lr@{}}
  \toprule
                       & \textbf{GTSRB}                              \\ \midrule
  Baseline            & \qty{1.1}{\second} \\
  DL2                 & \qty{1.1}{\second}      \\
  $T_\text{G}$     & \qty{1.1}{\second}        \\
  $T_\text{\L K}$ & \qty{1.1}{\second}       \\
  $T_\text{RC}$    & \qty{1.1}{\second}       \\
  $T_\text{YG}$    & \qty{1.2}{\second}       \\ \bottomrule
\end{tabular}
  \end{subtable}
\end{table}
The experiments were conducted on a system with a 4.5\,GHz AMD Ryzen 9 77950X and a GeForce RTX 4090.
The average epoch training times are shown in \Cref{tab:runtimes}.
In contrast to the original DL2 experiments~\cite{fischerDL2TrainingQuerying2019}, we did not observe significant overhead in our experiment.
We note that this is due to our reuse of the model output from the cross-entropy loss calculation in the forward pass for our logical loss calculation.
If our constraints required a separate forward pass, we would most likely see the same overhead as observed in the DL2 experiments.

\subsection{Plots of the Fuzzy Logic Implications}

\subsection{Plots of Prediction and Constraint Accuracy for Different Values of $\lambda$}

\newpage

\newpage

\end{document}